\documentclass[aps,prl,superscriptaddress,twocolumn,floatfix,citeautoscript]{revtex4-1}
\usepackage[pdftex]{graphicx}
\usepackage{color}
\usepackage{subfigure}
\graphicspath{{./figures/}} 
\graphicspath{{./pictures/}}

\newcommand{\ORNL}{Materials Science and Technology Division,
 Oak Ridge National Laboratory, Oak Ridge, Tennessee 37831-6138, USA}

\newcommand{\UTK}{Department of Materials Science, University of Tennessee, Knoxville,TN,USA}

\begin{document}

\title{Slow relaxation of cascade-induced defects in Fe}


\author{Laurent Karim B\'{e}land}
\author{Yuri N. Osetsky}
\author{Roger Stoller}
\affiliation{\ORNL}
\author{Haixuan Xu}
\email[]{xhx@utk.edu}
\affiliation{\UTK}

\date{\today}

\begin{abstract}

On-the-fly kinetic Monte Carlo (KMC) simulations are performed to investigate slow relaxation of non-equilibrium systems. Point defects induced by 25 keV cascades in $\alpha$-Fe are shown to lead to a characteristic time-evolution, described by the \emph{replenish and relax} mechanism. Then, we produce an atomistically-based assessment of models proposed to explain the slow structural relaxation by focusing on the aggregation of 50 vacancies and 25 self-interstital atoms (SIA) in 10-lattice-parameter $\alpha$-Fe boxes, two processes that are closely related to cascade annealing and exhibit similar time signature. Four atomistic effects explain the timescales involved in the evolution: defect concentration heterogeneities, concentration-enhanced mobility, cluster-size dependent bond energies and defect-induced pressure. These findings suggest that the two main classes of models to explain slow structural relaxation, the Eyring model and the Gibbs model, both play a role to limit the rate of relaxation of these simple point-defect systems.

\end{abstract}


\maketitle




Many off-equilibrium physical systems exhibit a slow structural
relaxation toward their ground state. Examples include glasses \cite{knoll2009relaxation,amir2012relaxations}, colloids \cite{sperl2003logarithmic}, concrete \cite{vandamme2009nanogranular} and amorphous solids \cite{tsiok1998logarithmic,trachenko2007logarithmic}. The degree of relaxation of these systems, e.g. their potential energy, is a nearly linear function of the logarithm of time. For simplicity, the term logarithmic relaxation is used to describe this behavior.

Most models describe such aging as a sequence of activated processes that permit relaxation. As the system relaxes, the energy barriers of these processes increase , which delays aging by growing orders of magnitude in time. The literature contains a large number of such propositions. For example, in the Eyring model \cite{eyring1936viscosity}, barriers are linked to the energy recovered during relaxation through their coupling to shear strain. This idea, where the degree of relaxation has an effect on the height of the barriers, has been adapted and modified to include hierarchically constrained dynamics \cite{brey2001slow}, stress relaxation in the Burridge-Knopoff model \cite{huisman2006logarithmic}, models with a stress-induced barrier increase \cite{trachenko2007logarithmic,trachenko2007slow} and glassy polymer relaxation \cite{knoll2009relaxation,knoll2009knoll} after nano-indentation. On the other hand, the Gibbs model \cite{gibbs1983activation} and its variants \cite{crandall1991defect,roura2009comment,roura2009structural} argue that the system initially possesses a distribution of relaxation events with a near-constant density as a function of activation barrier, or rates described by a multiplicative stochastic process \cite{amir2012relaxations}, which leads to logarithmic relaxation. A newly proposed model links logarithmic time-evolution to the system moving from one local state to another, where the waiting time of each state is defined by a power law and where all states evolve simultaneously \cite{lomholt2013microscopic}. Due to a lack of atomistic evidence, the validity of many of these proposed models remains ambiguous and the identity of the drivers to logarithmic relaxation remains elusive.

Recently, we proposed a novel description of logarithmic relaxation, coined \emph{replenish and relax} \cite{beland2013replenish}, which was based on nanocalorimetric measurements combined with the kinetic Activation Relaxation Technique (k-ART) \cite{el2008kinetic,beland2011kinetic} simulations of the annealing of ion-implanted \emph{c}-Si. In this slightly disordered system, we showed that relaxation is caused by the aggregation, reconfiguration and annihilation of small point-defect clusters \cite{beland2013long}. 

Isothermal magnetic relaxation measurements of neutron-irradiated Fe also exhibit such long-time relaxation \cite{blythe1973magnetic}. Work concerning a related problem, the aggregation of vacancies in $\alpha$-Fe \cite{brommer2012comment,xu2013cascade,BrommerArxiv,GowonouArxiv} showed that the transition from a state involving mostly mono vacancies to a state with large clusters involves logarithmic decay that is well described by the \emph{replenish and relax} model. 

These simulations in Fe and Si provided an accurate description of the microscopic processes, but did not explain why the barriers to unlock relaxation were increasing.

In this letter, we provide novel evidence that logarithmic decay appears in off-equilibrium, albeit largely crystalline, systems through the interaction of point defects. Firstly, we show that the annealing of 25 keV cascades in $\alpha$-Fe leads to such relaxation over seven orders of magnitude in time. To complete the picture drawn by vacancy clustering in Fe and cascade annealing, which involves both vacancies and interstitials, we also present simulations of self-interstitial atom (SIA) clustering, that exhibit near-logarithmic behavior over 19 orders of magnitude in time. Secondly, we verify that these systems are described by \emph{replenish and relax}. Thirdly, we examine the drivers of this behavior in the simplest cases, vacancy and interstitial clustering, and identify four mechanisms that explain the gradual increase of barriers during relaxation: defect concentration heterogeneities, concentration-enhanced mobility, cluster-size dependent bond energies and defect-induced pressure. We determine if these mechanisms are related to the Eyring or to the Gibbs model.

Molecular dynamics (MD) was used to simulate a 25 keV cascade in Fe. After 15 ps, we extracted the atomic positions of 250000 atoms from the MD and used them as the starting point for on-the-fly KMC simulations at 650K. For these calculations, we use the Self-Evolving Kinetic Monte Carlo (SEAKMC) \cite{xu2011simulating,xu2012self}, an off-lattice, on-the-fly kinetic Monte Carlo method that has been shown to accurately simulate the time-evolution of defects in Fe, such as the aggregation of vacancies \cite{xu2013cascade}, as well as collisions and transformations of interstitial loops \cite{xu2013solving}.

To simulate the aggregation of SIAs, we replace 25 lattice-sites by $<110>$ dumbells in a 2000-atom Fe crystal, for a total of 2025 atoms. The system is simulated at 100 K for times reaching 100 seconds to 10 million seconds, using the Marinica2007 (M07)  \cite{malerba2010comparison} potential. We choose a low temperature in order to distinguish the various low-barrier (less than 0.3 eV) processes that lead to interstitial-cluster formation. For this task, we choose the k-ART \cite{el2008kinetic,beland2011kinetic,mousseau2012activation,joly2012optimization}, an off-lattice, self-learning, on-the-fly kinetic Monte Carlo that can exactly handle elastic effects and time-evolution of states interconnected by small-barriers. This algorithm was successfully used to investigate the time-evolution of point defects in \emph{c}-Si \cite{beland2011kinetic,beland2013replenish,beland2013long}, \emph{a}-Si \cite{beland2011kinetic,joly2013contribution}, SiC \cite{jiang2014accelerated} and Fe \cite{GowonouArxiv,beland2011kinetic,brommer2012comment,BrommerArxiv}. It combines kinetic Monte-Carlo rules \cite{bortz1975new} to NAUTY \cite{mckay1981practical,mckay2007nauty}, a topological analysis software package, and the Activation-Relaxation Technique \emph{nouveau} \cite{barkema1996event,malek2000dynamics,machado2011optimized,mousseau2012activation,beland2014strain}.

\begin{figure}
	\centering
    \includegraphics[width=9cm]{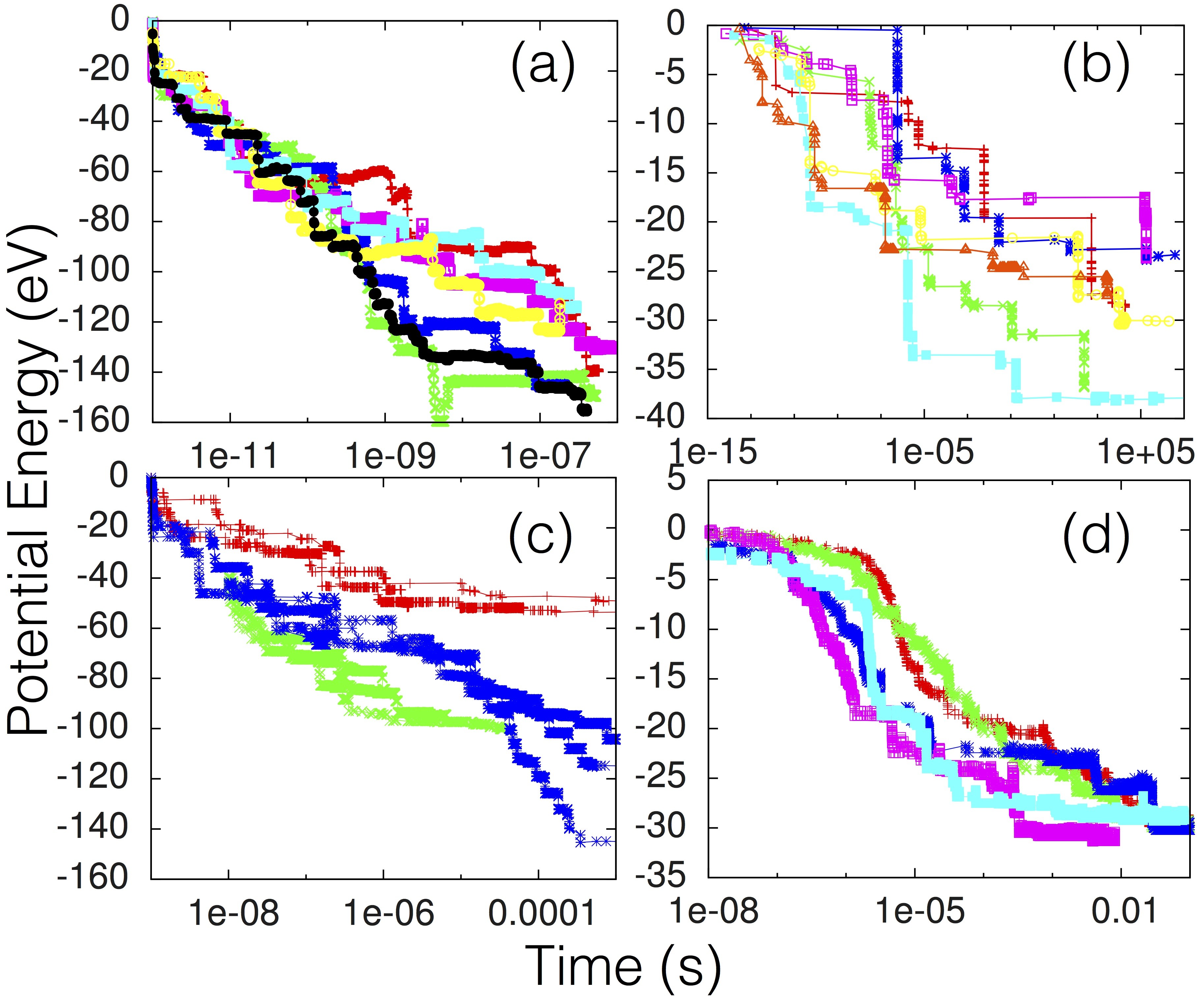}
    \caption{Time evolution of potential energy of four systems: 25 keV cascade annealing in Fe (a), 25-SIA aggregation in Fe (b), 3 keV \emph{c}-Si ion implantation annealing (c), and 50-vacancy aggregation in Fe (d).} 
\label{fig:runs} 
\end{figure}

The time-evolution for the 25 keV cascade annealing and SIA clustering is plotted in Fig. \ref{fig:runs}. We also show, for reference, the time-evolution of 50-vacancies aggregating and the annealing of 3 keV ion-implanted \emph{c}-Si. In all cases, we see that the potential energy relaxation takes place over logarithmic timescales. For reference, we also performed SIA-cluster aggregation runs with the A04 potential; the results were similar to those with the M07 potential, as shown in the Supplemental Material\cite{SupplMat}. The variation between each run for the SIA-clustering system is due to each simulation starting from a different random initial configuration. In the Supplemental Material \cite{SupplMat}, we show that each of these runs is roughly logarithmic.


\begin{figure}
	\centering
   \includegraphics[width=9cm]{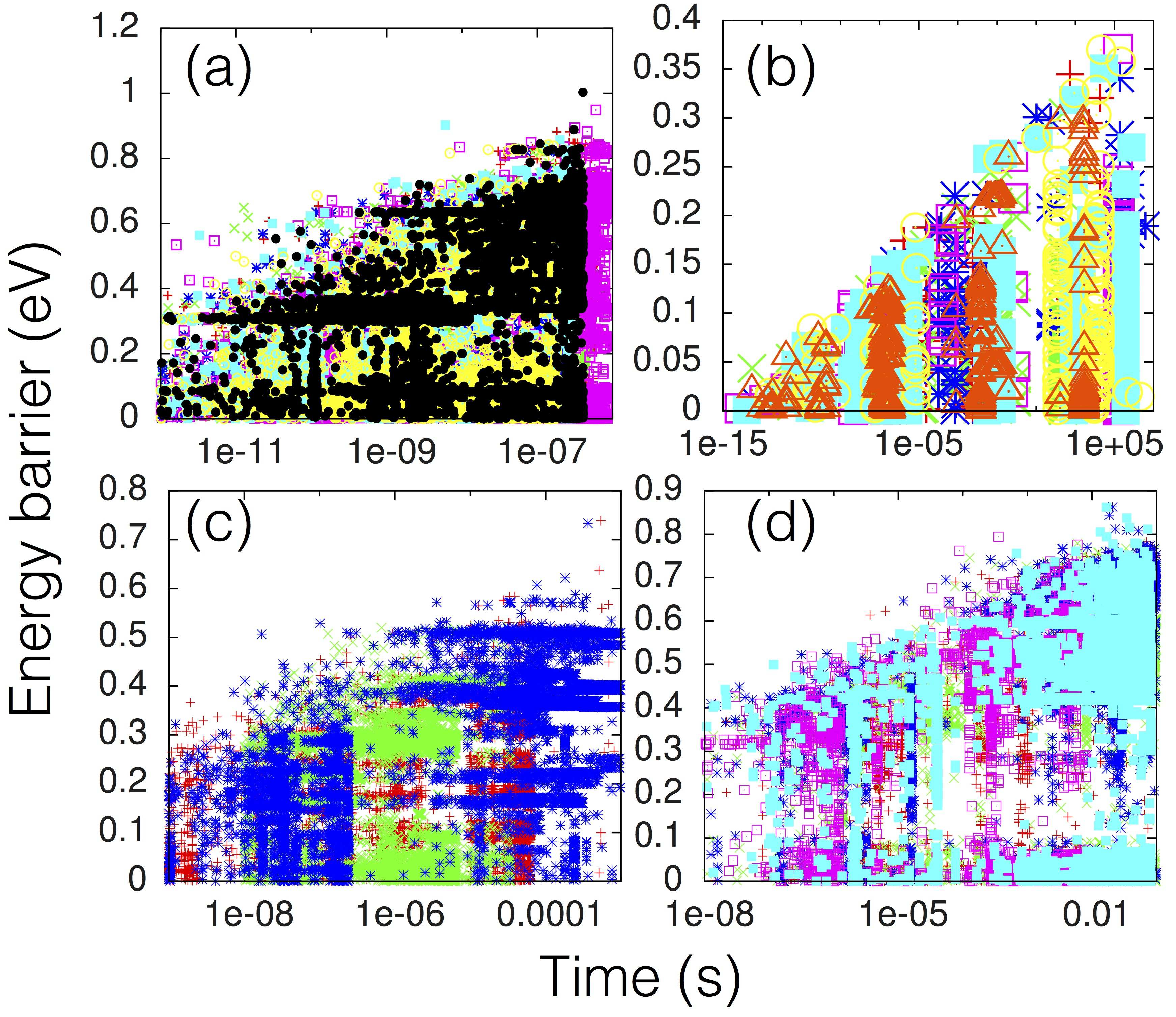}
    \caption{The activation barriers of the events executed during the simulations: 25 keV cascade annealing in Fe (a), 25-SIA aggregation in Fe (b), 3 keV \emph{c}-Si ion implantation annealing (c), and 50-vacancy aggregation in Fe (d).} 
\label{fig:barriers} 
\end{figure}

\begin{figure}
	\centering
    \includegraphics[width=9cm]{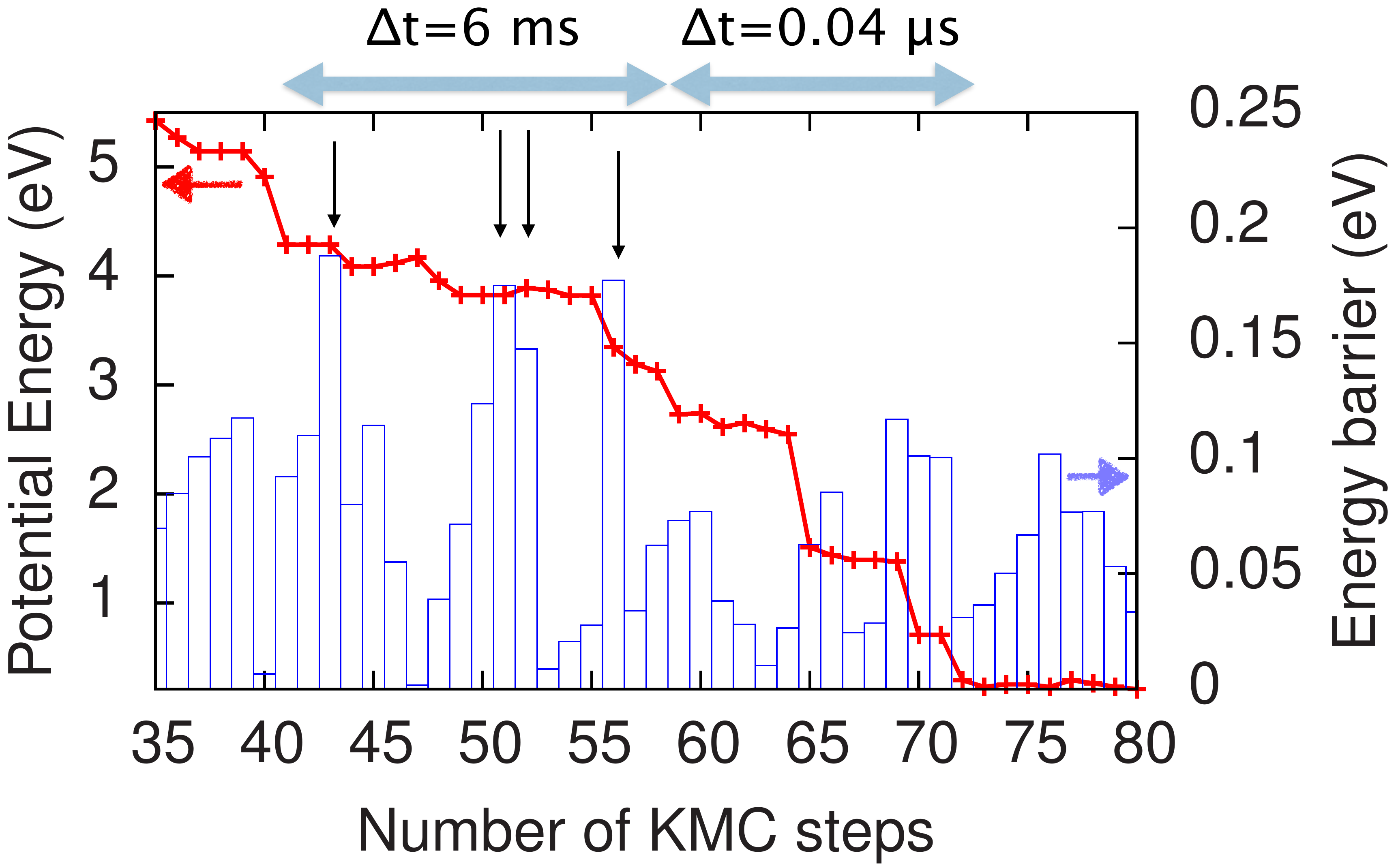}
    \caption{A representative example of the relaxation kinetics, taken from a 25-SIA run. The black arrows indicate bottleneck events, which account for most of the time-evolution. The blue bars represent the potential energy barrier of each event in the sequence. The red line represents the potential energy of the initial state for each event. The potential energy is shifted relative to that plotted in Fig. \ref{fig:runs} for simplicity.} 
\label{fig:example} 
\end{figure}

We plot the activation barriers that were crossed during each of these runs in Fig. \ref{fig:barriers} . The maximum energy-barrier increases exponentially with time, as expected from Poisson processes, and a wide distribution of barriers is executed in each time-frame. Combined to the fact that potential energy relaxation takes place on a linear scale relative to the number of KMC steps (see Fig. \ref{fig:example} and the Supplemental Material \cite{SupplMat}), this indicates that the largest barriers in each time-frame act as a bottleneck. 

In Fig. \ref{fig:example}, we show a typical relaxation sequence. This example is taken from one of the 25-SIA runs. We see two plateaux in potential energy: from KMC step 41 to step 55 and from step 72 to step 80. We also see two rapid decreases in energy (i.e. relaxation phases): from step 25 to step 41 and from step 55 to step 72. The first plateau is characterized by the presence of four large activation barriers, that act as a bottleneck for time-evolution. In contrast, the phases with rapid energy relaxation were executed without crossing large activation barriers. From a time-evolution perspective, the system took 6 ms to execute events in the first plateau, while it took only 40 ns to execute events in the second relaxation phase. Generally, the events with large activation barriers do not directly lead to lower-energy states. Instead, they give access to a new region of the potential energy surface were such lower-energy states are accessible by crossing small energy-barriers. In the Supplemental Material \cite{SupplMat}, we provide data characterizing all the bottleneck-events and all the events leading to lower-energy states in our runs, which agrees with the mechanisms illustrated in Fig. \ref{fig:example}. 


As a whole Fig. \ref{fig:runs} through Fig. \ref{fig:example} show that all these systems relax logarithmically through the \emph{replenish and relax} mechanism. Indeed, the energy relaxation of the system progressively decelerates due to growing potential energy barriers that need to be crossed in order to access a section of the energy landscape where relaxation events are present. This distribution of relaxation events is emptied and the potential energy no longer decreases, until an event with a large barrier is executed and replenishes it.

It is remarkable that these four different systems all lead to relaxation on logarithmic timescales, described by  \emph{replenish and relax}. Indeed, this indicates that such long-time structural evolution is not necessarily associated to complex materials, i.e. amorphous solids, colloids, glasses, polymers, but can be driven by the interaction of simple point defects in a pure cristal (all our configurations had more that 98\% cristalline atoms). We also note that simulations in \emph{a}-Si indicate logarithmic relaxation may appear after ion-implantation \cite{pothier2011flowing,joly2013etude}. This apparently general behavior also raises the question as to the drivers of this type of annealing. Our simulations give a fully atomistic account of relaxation on logarithmic timescales over several orders of magnitude in time and should thus provide a much clearer picture, with less room for interpretation and speculation than previous studies. We focus on the Fe 50-vacancies and Fe 25-SIA cases, as the absence of recombination of point defects greatly simplifies the analysis. 

\begin{figure}
	\centering
    \includegraphics[height=3.7cm]{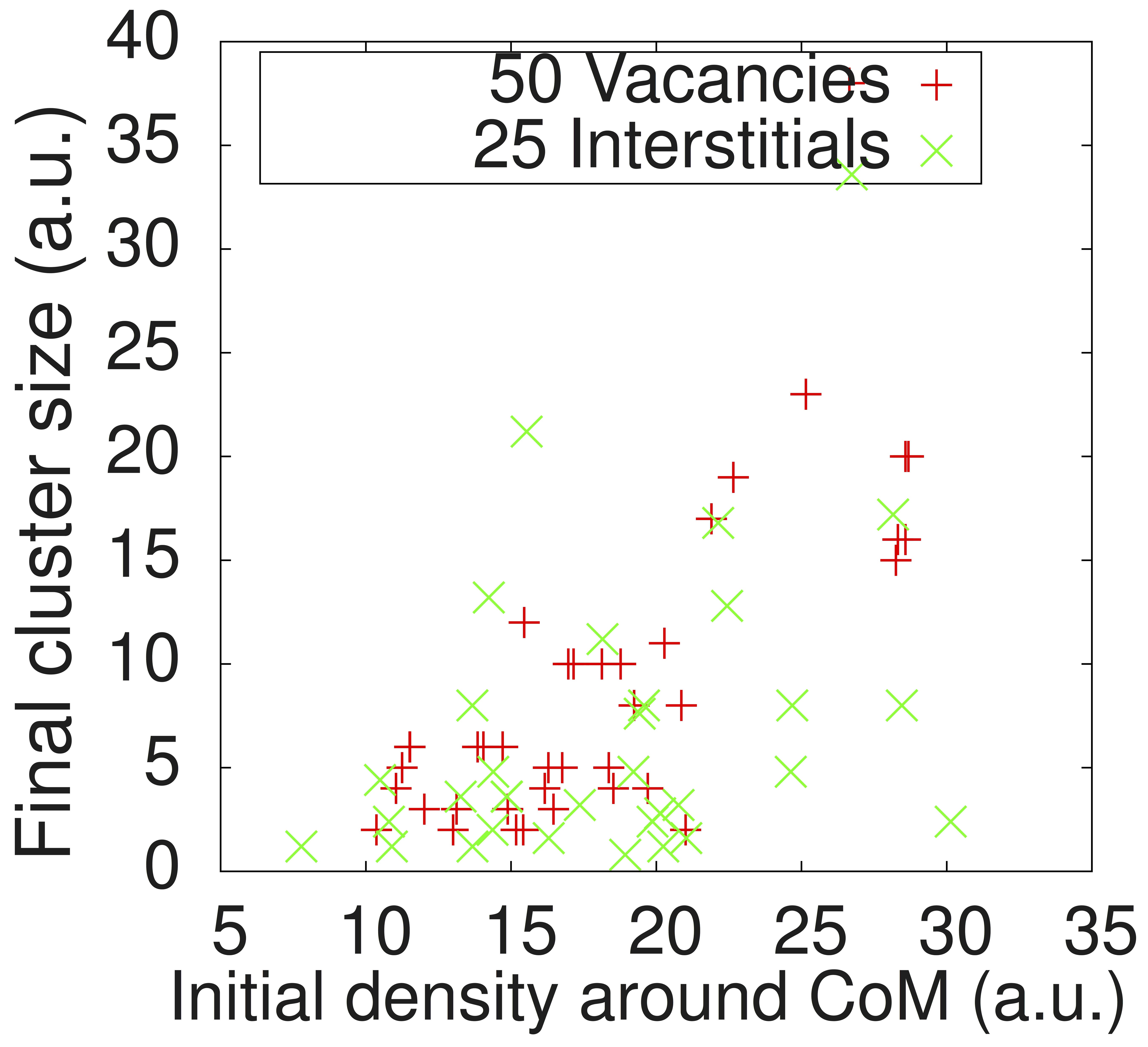}
    \includegraphics[height=3.7cm]{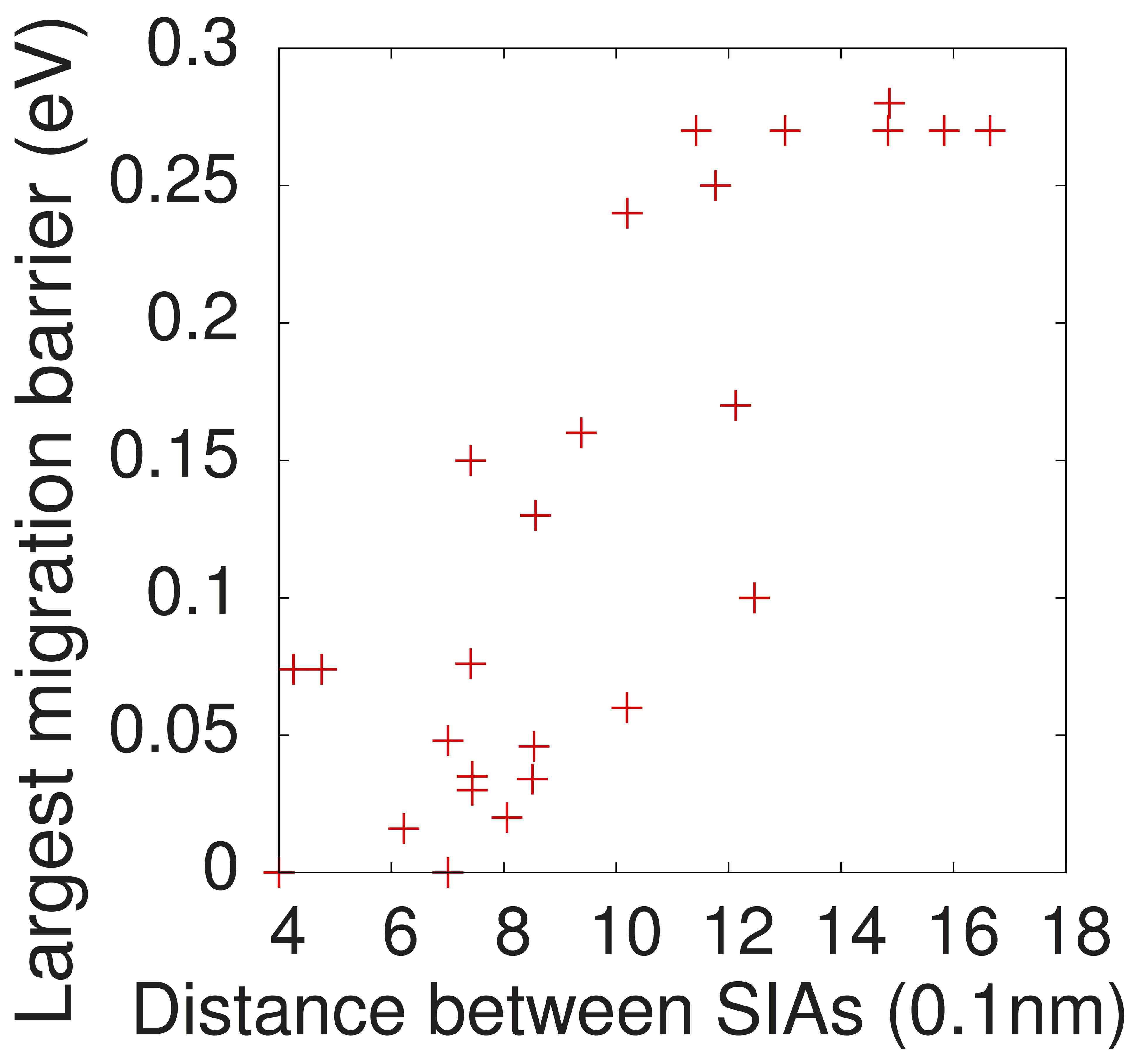}
    \caption{Left panel: the size of the clusters in the final configuration after the 50-vacancy ($R^2$=0.58) and 25-SIA ($R^2$=0.14)  simulations as a function of the initial density of point defects in the neighborhood of their CoM. Right panel: The largest barrier crossed when two SIAs set at a given distance from each other aggregate ($R^2$=0.69).} 
\label{fig:clus_dens} 
\end{figure}

We identified four elements that lead to relaxation over logarithmic timescales:

1- Vacancies are typically limited to one or two jumps before aggregating, which promotes large cluster formation in high-density areas. Because SIA-clusters can execute long jumps, this effect is much weaker.  Fig. \ref{fig:clus_dens} quantifies this effect, which is in aggreement with visual inspection (see e.g. the movies in the Supplemental Material \cite{SupplMat}). We identified the size and the center of mass (CoM) of each defect cluster at the end of each simulation and calculated the density of defects that were present in the neighborhood of the CoM in the local initial configuration, using gaussian kernel averaging. In the case of the 50-vacancy system, 57 \% of the size of the final clustered can be explained by the initial density of defects, while it explains less than 20 \% in the 25-SIA system. We infer that the heterogeneity in the initial density of defects leads to a distribution of clusters of different sizes, which each have their own distribution of barriers to \emph{replenish} the potential energy landscape. On aggregate, this will lead to a wide distribution of such barriers. This is in general agreement with the Gibbs model and its variations. This effect seems important in the vacancy case, but weak in the interstitial case.

2- In the first stages of relaxation, the systems contains an important density of mobile point defects and point defect clusters, which decreases as the system evolves. In Fig. \ref{fig:barriers}, we observe that relaxation occurs with barriers lower than that of diffusion of isolated point defects and isolated point defect clusters. This value this corresponds to 0.3 eV for SIAs and 0.64 eV for mono-vacancies \cite{malerba2010comparison} (di- and tri-vacancies are know to diffuse slightly faster). This indicates that mobility is enhanced in the first stages of relaxation. For vacancies, it is known that interactions between nearby vacancies can lead to low-barrier vacancy jumps and that this effect is stronger when distances are shorter \cite{kapinos1991effect,BrommerArxiv}. In Fig. \ref{fig:clus_dens}, we show that high local concentration of SIAs also leads to high mobility. In other words, migration is enhanced by defect proximity and concentration. During relaxation, as the cluster size increases and potential energy decreases, the distance between mobile point defects increases and activation barriers increase. This is a mechanism in general agreement with Eyring-like models.

3- Diffusivity of large vacancy clusters (tetra-vacancies and larger) decreases with cluster size and thermal lifetimes increase with cluster size \cite{BrommerArxiv}. Thus, as the clusters grow in size, one would expect the overall relaxation-limiting barriers to increase. This mechanism observed for the vacancy case is in general agreement with Eyring-like models. 

In the case of SIA-clusters, the relationship is not well defined. SIAs bind in a large number of metastable sessile and glissile configurations \cite{marinica2011energy,beland2011kinetic}. The activation barriers that lead to reconfiguration and interactions with nearby defects are unpredictable and largely independent of cluster size \cite{terentyev2007dimensionality,anento2010atomistic}, especially since the configurations are generated through off-equilibrium dynamics. Both visual inspection of simulations and the data presented in the left panel of Fig. \ref{fig:clus_dens} indicate that mobility is weakly, if at all, related to cluster size during these simulations.

\begin{figure}
	\centering
    \includegraphics[height=3.7cm]{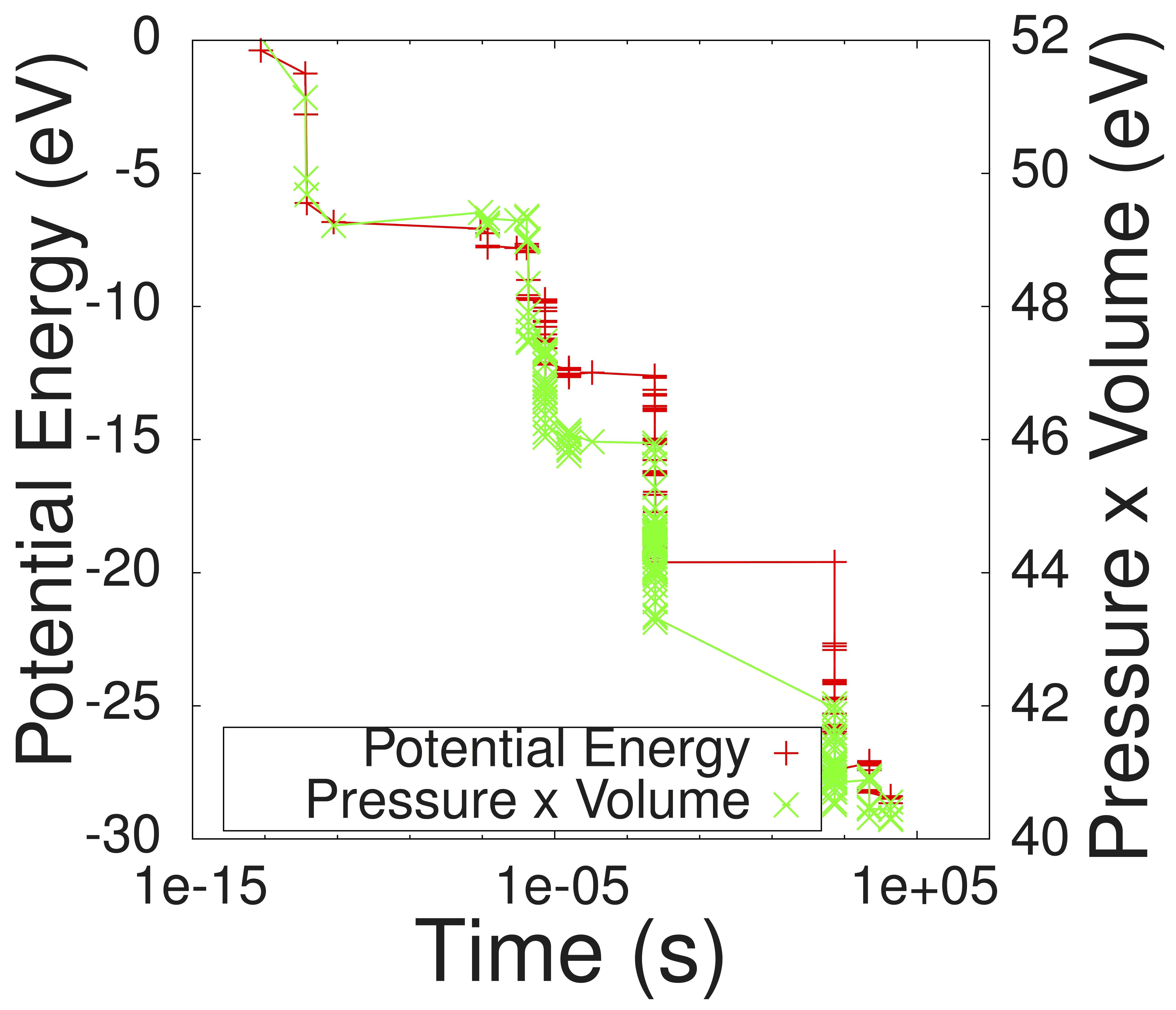}
    \includegraphics[height=3.7cm]{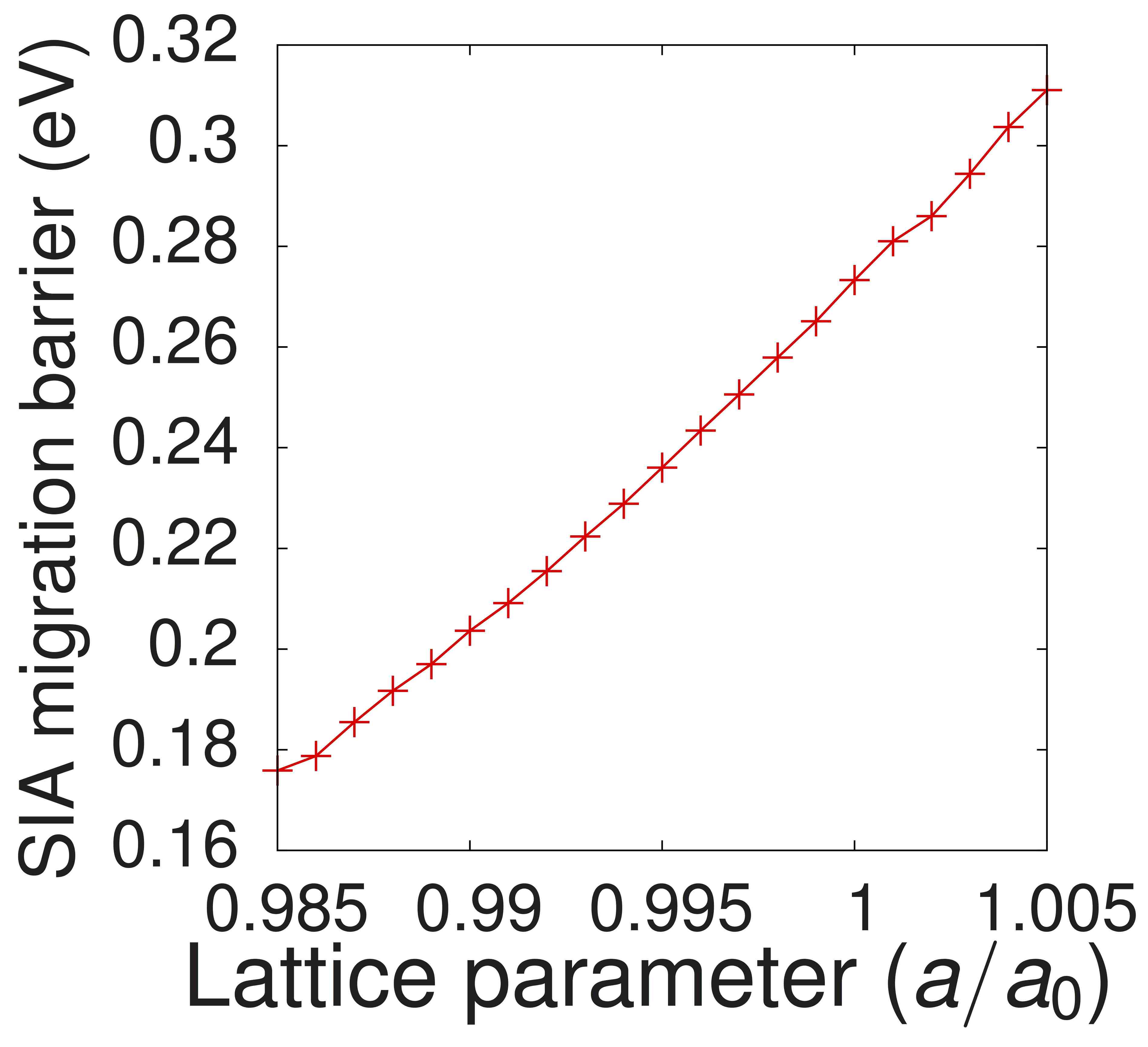}
    \caption{Left panel: the hydrostatic pressure of the 25-SIA system and its potential energy as a function of time. Right panel: the effect of hydrostatic pressure on SIA mobility.} 
\label{fig:stress} 
\end{figure}

4- In our NVT runs, aggregation of point defects relieved pressure. In the left panel of Fig. \ref{fig:stress}, we show the typical evolution of the total pressure, computed as the sum of all the traces of the atomic virial stress tensors, for the 25-SIA system. The aggregation of SIAs leads to a large decrease in hydrostatic stress, which follows the relaxation in potential energy very closely. Shear stresses, including the Von Mises stress, are an order of magnitude smaller and, moreover, do not have a systematic trend during relaxation. We assessed the impact of pressure on SIA mobility by measuring the activation barrier for a SIA in a box with various lattice-parameters. The results are plotted in the right panel of Fig. \ref{fig:stress}. We see that 1\% compressive hydrostatic strain leads to a 26\% decrease in barrier energy for the M07 potential. Thus, when the system is unrelaxed, mobility is enhanced through pressure. As the system relaxes, this effect decreases, causing the barriers to increase towards their unstrained value. We note that in the A04 potential, the effect of pressure on mobility is much smaller, a 1\% compressive hydrostatic strain leads to a 7\% decrease of the activation barrier. 

For mono-vacancies, the effect of stress on diffusion is small. A 1\% tensile hydrostatic strain leads to a 6\% increase of the activation energy, both in the A04 and M07 potential, which means that the tensile strain release during relaxation accelerates time-evolution, albeit by a small quantity compared to the other effects. 

In other words, the interplay between the overall pressure in the system, the state of relaxation and mobility plays an important role to logarithmically limit the time-evolution of SIA aggregation as described by the M07 potential, in general agreement with Eyring-like models, but plays a much smaller role if this process is described with the A04 potential. In the vacancy case, the effect is very small and in fact plays against logarithmic relaxation. 

We also looked at variations in pressure during events corresponding to bottlenecks. We saw no clear relationship between the pressure change from the initial state to the saddle point and the activation barrier, which indicates that the model of Ref. \cite{trachenko2007slow} does not apply to these simple point-defect systems.

To conclude, through the use of SEAKMC and k-ART simulations, we found that simple point-defect interactions can lead to relaxation on logarithmic timescales described by the \emph{replenish and relax} process. We showed three examples in Fe and one in \emph{c}-Si that fit this description. For the simplest cases, the aggregation of 50 vacancies in Fe and the aggregation of 25 interstitial in Fe, we identified four atomistic mechanisms explaining how the barriers to relaxation increase as the system relaxes: defect concentration heterogeneities, concentration-enhanced mobility, cluster-size dependent bond energies and defect-induced pressure. Some of these mechanisms support the Gibbs model of relaxation, while others give credence to variations of the Eyring model.

The variety of mechanisms involved in these seemingly simple problems is striking. It is also clear that slow relaxation is caused by events localized around interacting defects. Furthermore, this work provides a framework, based on accelerated atomistic simulations, to approach more complex problems, that may well share common mechanisms with point defect kinetics. This may not necessarily be the case, but the generality of our observations motivate such further investigation.

Research at the Oak Ridge National Laboratory sponsored by the U.S.
Department of Energy, Office of Basic Energy Sciences, Materials
Sciences and Engineering Division, "Center for Defect Physics", an
Energy Frontier Research Center. LKB acknowledges a fellowship awarded
by the Fonds Qu\'eb\'ecois de recherche Nature et Technologies.The k-ART software is available upon request to Normand Mousseau and SEAKMC is available upon request to the authors.

\bibliography{Slowbib}
 
\end{document}